\documentclass{article}
\usepackage{graphicx}
\def\be{\begin{equation}}
\def\ee{\end{equation}}
\def\bea{\begin{eqnarray}}
\def\eea{\end{eqnarray}}
\def\Journal#1#2#3#4{{#1} {\bf #2}, #3 (#4)}
\def\NC{\em Nuovo Cimento }
\def\PLB{{\em Phys. Lett.}  B}

\def\NCA{{\em Nuovo Cimento} A}
\def\LNC{\em Lett. Nuovo Cimento }
\def\RNC{\em Rivista Nuovo Cimento }
\def\PR{\em Phys. Rev. }
\def\PLB{{\em Phys. Lett.} B}
\def\EPJC{{Eur. Phys. Journ.} C}

\begin{document}
\begin{center}
{\bf \Large Soft Dipole Pomeron in hadronic elastic\\ and in deep
inelastic scattering }

\medskip
E. Martynov

\medskip
Bogoliubov Institute for Theoretical Physics, 03143, Kiev-143,
Ukraine
\\E-mail: martynov@bitp.kiev.ua
\vskip 0.5cm
\begin{minipage}{12.cm}
{\bf \large Abstracts.} A brief review on the Dipole Pomeron model is
given. The model not only describes data on hadron-hadron
interactions, but also allows to describe data on the proton
structure function with a $Q^2$ independent intercept. Moreover the
chosen Dipole Pomeron has an intercept equal to one and does not
violate unitarity limit on the total elastic cross-section.
\end{minipage}
\end{center}

\vskip 0.4cm
The Regge approach \cite{Coll} is one of the most
powerful method to investigate hadronic interactions at high
energies. In spite of many unsolved problems, its actual application
is slowed down because of lacking new experimental data. Now, the
main efforts of the experimentalists and theoreticians are
concentrated on the deep inelastic scattering (DIS) and related
processes. Again the Regge method shows its effectiveness in a wide
kinematical domain.

It follows quite evidently from the observed growth of the hadronic
total cross-sections and of the structure functions that the true
Pomeron singularity is more complicated than that a simple $j$-pole. Even
if a simple pole with an intercept larger than unity is used as an input,
it should be unitarized in order to restore the Froissart-Martin
bound \cite{F-M}, $\sigma_t(s)<C\ln^2s$. Unfortunately a
well defined strict procedure of unitarization is still absent, while
approximate methods are used; as consequence, we do not know what
is really the true Pomeron.

The well known and quite popular example of a model with an input
Pomeron violating the unitarity bound (because it has $\alpha_{\cal
P}(0)\approx 1.08>1)$) is the Donnachie-Landshoff Pomeron (D-L)
\cite{DL1}. Moreover, and regrettably, in some papers this model is
identified with the Regge theory and its inability to describe some
experimental data is declared as a problem of Regge theory. However,
and happily, the D-L Pomeron is not the only phenomenological
possibility to describe the above mentioned growth of observed
quantities. Another way to satisfy this property is to suppose that
the Pomeron is harder than a simple pole at $j=1$. The simplest
realization of such an hypothesis is the Dipole Pomeron model
\cite{DPrev}.

The Dipole Pomeron (DP) is a combination of the double and simple
poles in the partial $j$-plane amplitude
$$
\Phi(j,t)\propto \frac{\phi(j,t)}{[j-\alpha_{\cal P}(t)]^2}\approx
\frac{\phi_1(t)}{[j-\alpha_{\cal P}(t)]^2}+
\frac{\phi_2(t)}{j-\alpha_{\cal P}(t)}
$$
 with the trajectory
$
\alpha_{\cal P}(t)\approx 1+\alpha'_{\cal P}t$, linear at small $t$.
The Dipole Pomeron is the unitarity limit for an isolated $j$-singularity
with a unit intercept and with a linear trajectory at small
$t$. If
$$
\Phi(j,t)\propto (j-1-\alpha't)^{-\nu -1}\quad \mbox{then}\quad
\sigma_{tot}\propto \ln^{\nu}s/s_0, \quad \sigma_{el}\propto
\ln^{2\nu-1}s/s_0,
$$
$$
\sigma_{el}/\sigma_{tot}\leq 1 \quad \Rightarrow \quad \nu \leq 1.
$$
Thus in the present Dipole Pomeron model at $s\to \infty$:
$$
\sigma_{tot}^{(\cal P)}(s)\propto \sigma_{el}^{(\cal P)}(s)\propto
\sigma_{inel}^{(\cal P)}(s)\propto \ln(s/s_0).
$$
\vskip 0.4cm
\noindent
{\bf \large The Dipole Pomeron in hadronic
processes.} At preasymptotic energies the contribution of other
reggeons ($f, \omega $ {\it etc...}) should be added. If only $pp$
and $\bar pp$ amplitudes are considered one can write \footnote{We do
not consider here Odderon because it is out of the subject of talk.}
\begin{equation}\label{1}
A^{(\bar pp)}_{(pp)}(s,t)={\cal P}(s,t)+f(s,t)\pm \omega (s,t)
\end{equation}
with
\bea
{\cal
P}(s,t)&=&i[g_1(t)\ln(-is/s_0)+g_2(t)](-is/s_0)^{\alpha_{\cal
P}(t)-1},\quad s_0=const,\\ r(s,t)&=&\eta_r\,
g_r(t)(-is/s_0)^{\alpha_r(t)-1}, \qquad r=f,\omega, \quad
\eta_f=i,\quad \eta_\omega =1.
\eea

The DP model was applied \cite{DGLM} to the analysis of available data on the
total cross-sections and on the real to imaginary part ratios (at $t=0$ and
$\sqrt{s}\geq 5$ GeV) for meson-nucleon and nucleon-nucleon
interactions. It leads to the best description ($\chi^2/d.o.f.\approx
1.12$) comparing with other models. A new very detailed
and circumstantial analysis of these data confirming our conclusions
is presented at this conference in the talk of V. Ezhela \cite{Ezh}

Furthermore, in the framework of a modified additive quark model
(where corrections for the coupling of Pomeron with two quark lines
as well as new counting rules for the secondary reggeons were taken
into account) \cite{DGMP} the Dipole Pomeron was used to describe
simultaneously cross-sec\-tions of $p^{\pm}p, \pi^{\pm}p, \gamma p,
\gamma \gamma $ interactions, related by factorization conditions.

An important property of the Dipole Pomeron model is that all fits
give a high value of the $f$-reggeon intercept, $\alpha_f(0)\approx
0.8\div 0.82$. Does such an intercept contradict to the data on the
$f$-trajectory known from a resonance region? The answer is "yes" if
trajectory is assumed to be linear. However besides of general
theoretical arguments in favor of nonlinear trajectories, the
experimental data on resonances lying on $f$-trajectory indicate its
nonlinearity (see Fig.1). A more realistic trajectory than those
shown in the Fig.1,
$
\alpha_f(t)=\alpha_f(0)+\gamma_1(\sqrt{4m_{\pi}^2}-\sqrt{4m_{\pi}^2-t}\
)+ \gamma_2(\sqrt{t_1}-\sqrt{t_1-t}\ ),
$
gives $0.77<\alpha_f(0)<0.87$.

In the Fig.2, we show how $\alpha_f(0)$ is correlated with a power of
$\ln s$ in behaviour of the total cross-section if the amplitudes are
parameterized in the form (1-3), but with the replacement
$\ln(-is/s_0)\rightarrow \ln^\gamma(-is/s_0)$ (see \cite{DGLM} for
details). \vskip 0.4cm
\begin{figure}[h]
\begin{minipage}[b]{7.cm}
\begin{center}
\includegraphics[scale=0.38]{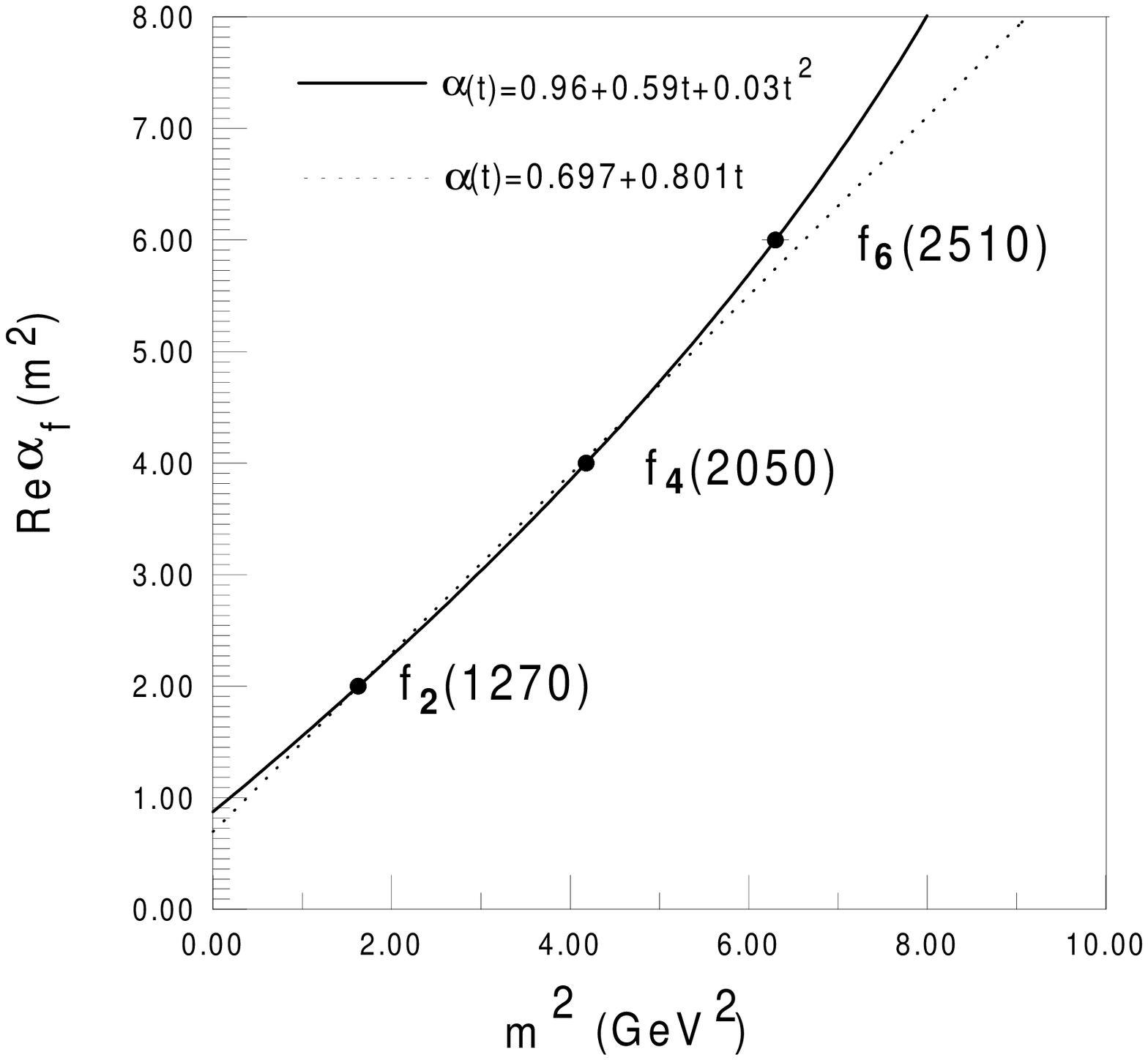}
\caption{Real part of $f$-trajectory} \label{fig:freal}
\end{center}
\end{minipage}
\hskip 1.cm
\begin{minipage}[b]{7.2cm}
\begin{center}
\includegraphics[scale=0.45]{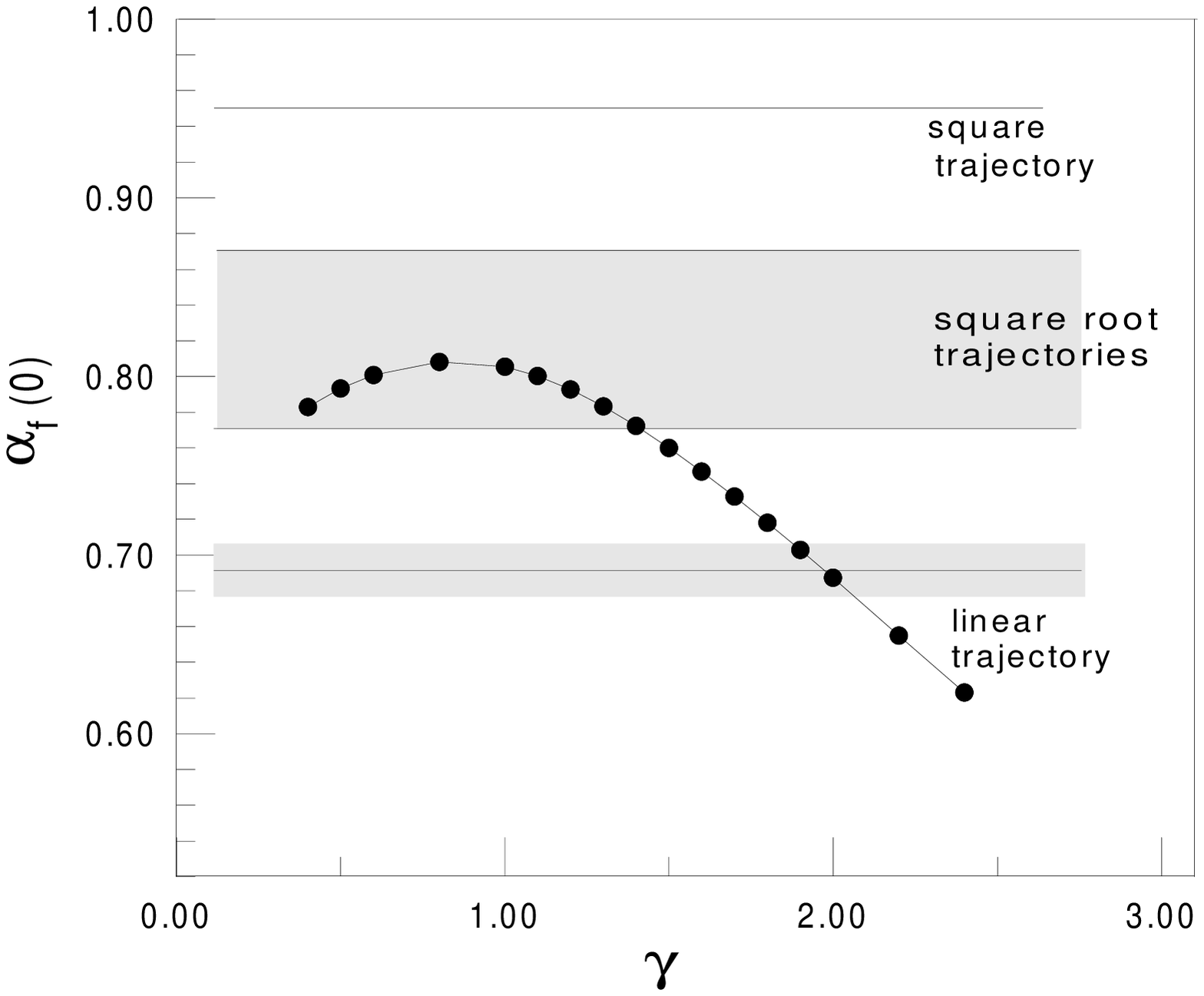}
\caption{Intercept of $f$-trajectory vs.
$\gamma$.\label{fig:af_vs_g}}
\end{center}
\end{minipage}
\end{figure}

\noindent {\bf \large The Dipole Pomeron in deep inelastic
scattering.} We support and investigate the point of view that there
is only one Pomeron (two Pomerons, "soft" and "hard", are considered
in \cite{Bert,DL2}). This unique Pomeron is universal and
factorizable (at least the main term should satisfy factorization at
c.m.s. energy $W\gg m_p$), {\it i.e.} it is the same in all
processes, only vertex functions depend on interacting particles.
Thus we think that Pomeron has a $Q^2$ independent trajectory. As in
the pure hadron case we use a Pomeron trajectory with a unit intecept
($\alpha_{\cal P}(0)=1$).

Defining the Dipole Pomeron model for DIS,
we start from the
expression connecting the transverse cross-section of $\gamma^*p$
interaction to the proton structure function $F_2^p$ and the optical
theorem for forward scattering amplitude \be\label{4}
\sigma_{T}^{\gamma^*p}=\Im m
A(W^2,Q^2;t=0)=\frac{4\pi^2\alpha}{Q^2(1-x)}(1+4m_p^2x^2/Q^2)
F_2^p(x,Q^2), \ee
where $\sigma_L^{\gamma^*p}=0$ is assumed. The
forward scattering at $W^2=Q^2(1/x-1)+m_p^2$ far from the threshold
$W_{th}=m_p$ is dominated by the Pomeron and the $f$-reggeon
 \footnote{We ignored an $a_2$-reggeon considering the
$f$-term as an effective one at $W>3$ GeV.}
$$
\begin{array}{c}
A(W^2,Q^2;t=0)=P(W,Q^2)+f(W,Q^2), \qquad P(W^2,Q^2)=P_1+P_2,
 \\ P_1=iG_1(Q^2)\ln(-iW^2/m_p^2)(1-x)^{B_1}, \qquad
P_2=iG_2(Q^2)(1-x)^{B_2}. \\
f(W^2,Q^2)=iG_f(Q^2)(-iW^2/m_p^2)^{\alpha_f(0)-1}(1-x)^{B_f}.
\end{array}
$$
\begin{figure}[ht]
\begin{minipage}[b]{6.5cm}
It evidently follows from the experimental data (Fig.3) that
$Q^2\sigma^{\gamma^*p}$ decreases with $Q^2$ at least at high $Q^2$.
Therefore we choose $$G_i(Q^2)=g_i/(1+Q^2/Q_i^2)^{D_i},$$ expecting
$D_i>1$ at high $Q^2$. Here we only note that as it follows from the
fit, $D_i$ and $B_i$ should be functions of $Q^2$. By hypothesis each
of them varies between two constants when $Q^2$ goes from $0$ to
$\infty $. The details of the parameterization of the real functions
$D_i(Q^2), B_i(Q^2)$ (as well as the references for the data) are
given in the paper \cite{DLM}.
\end{minipage}
\hskip 1.5cm
\begin{minipage}[b]{6.cm}
\includegraphics[scale=0.33]{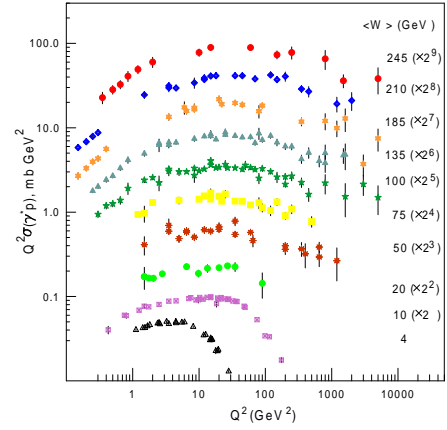}
\caption{Data on $Q^2\sigma^{\gamma^*p}$ as function of $Q^2$ at
fixed $W$. \label{fig:qsig}}
\end{minipage}
\end{figure}

A fit was performed to the 1209 experimental points in the region
$$
W\geq 3 GeV^2,\quad 0\leq Q^2\leq 5000 GeV^2,\quad 0\leq x\leq 0.85,
$$
and a quite good description of data was obtained: $\chi^2/d.o.f.=1.11$
and $\chi^2/d.o.f.=0.77$ if $x\leq 0.1$ and any dependence on $x$ is
canceled ($B_i=0$).

A detailed analysis \cite{DLM} of the fit leads to the following
conclusions:

$\bullet$ The possibility to describe available data with the
Pomeron that does not violate the Froissart-Martin limit is shown.

$\bullet$ The preasymptotic contributions, $P_2$ and $f$, play an
important role in the considered kinematical domain.

$\bullet$ The Dipole Pomeron has a $Q^2$ independent intercept. There
is no contradictions with the "experimental data" on
$\Delta(Q^2)=\alpha_{\cal P}-1$.
\begin{figure}[ht]
\begin{minipage}[b]{7.7cm}
Indeed, the effective Pomeron intercept is extracted from the data on
$F_2^p(x,Q^2)$ making use of the simplified parameterization of the
structure function $F_2^p=a+bx^{-\Delta} \quad \mbox{at fixed} \quad
Q^2.$ On the second hand, one can write
$$F_2^p(x,Q^2)=G(Q^2)(1/x)^{\Delta_{eff}(x,Q^2)}$$
and consequently
$$
\frac{\partial \ln F_2^p}{\partial \ln
(1/x)}=\Delta_{eff}+x\frac{\partial \Delta_{eff}}{\partial x}\ln
(1/x)
$$
If $\Delta_{eff}(x,Q^2)\approx {\rm const} $ at $x\ll 1$ then
$$
\frac{\partial \ln F_2^p}{\partial \ln (1/x)}\approx \Delta_{eff}.
$$
Comparison of this derivative in our model with the experimental data
on $\Delta_{eff}$ (or $\lambda_{eff}$ in the notation of the
experimentalists) is shown in Fig.4. The open triangles and the curve
correspond to the $\frac{\partial \ln F_2^p}{\partial \ln (1/x)}$
calculated in the given model.
\end{minipage}
\hskip 1.cm
\begin{minipage}[b]{6.6cm}
\includegraphics[scale=0.35]{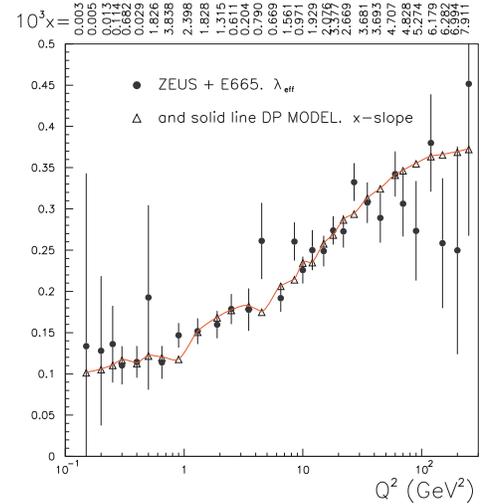}
\caption{$\Delta_{eff}(Q^2)$, extracted from experimental data and
compared to the $x$-slope calculated in the Dipole Pomeron Model.
\label{fig:effinterc}}
\end{minipage}
\end{figure}

There is an important prediction of the model concerning
$\frac{\partial \ln F_2^p}{\partial \ln (1/x)}$ at fixed $x$. This
$x$-slope begins to decrease when $Q^2>$ $\sim 100GeV^2.$ It would be
 interesting to compare the model with new high $Q^2$ data when they
are published. A comparison of this quantity in our model \cite{DLM},
in a QCD motivated ALLM model \cite{ALLM} and in the interpolating
model \cite{LKP} (between a Regge-type behaviour and a solution of
the evolution equation) is shown in Fig.5
\vskip 0.3cm
\begin{figure}[ht]
\begin{center}
\includegraphics[scale=0.7]{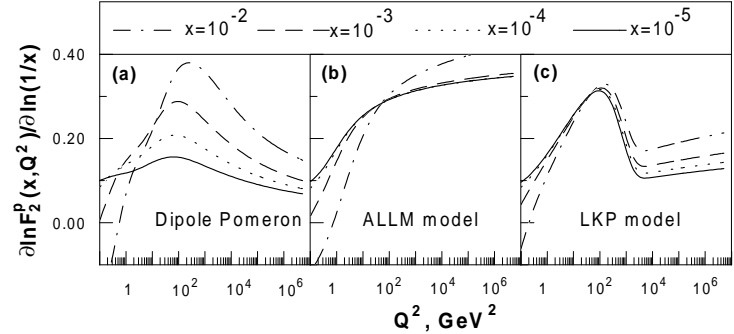}
\caption{A behaviour of $x$-slope in three models (see text for
details). \label{fig:x_slope}}
\end{center}
\end{figure}

The Soft Dipole Pomeron model well describes also the $Q$-slope or
$\frac{\partial F_2^p(x,Q^2)}{\partial \ln Q^2}$. A more detailed
discussion of the properties of the Dipole Pomeron Model (comparison
with the data on $\sigma^{\gamma^*p}, F_2^p$ and so on) as well as
some predictions of the model are given in the paper \cite{DLM}.
\vskip 0.4cm
\noindent
{\large \bf Conclusion.} The available data on
hadron-hadron and photon-hadron cross-sections can be described in a
traditional Regge approach with a Soft Pomeron. It is not necessary
to use an intercept depending on $Q^2$ and higher than one. A scale
in $Q^2$, where a nonperturbative Regge behaviour is transformed to a
QCD perturbative one, is still unknown in the presented approach and
requires an additional consideration.

I thank very much my coworkers P.Desgrolard, M.Giffon, A.Lengyel and
E.Predazzi for many fruitful discussions, for interesting and useful
collaboration.

\end{document}